\documentstyle[12pt]{article}
\advance\textheight by \topskip
\begin{document}
\begin{titlepage}
\begin{flushright}
UH-IfA-94/35\\
SU-ITP-94-13\\
YITP/U-94-15\\
hep-th/9405187\\
\end{flushright}
\vskip  2 truecm
\begin{center}
 {\Large\bf   REHEATING AFTER INFLATION}
\vskip 1.6cm
{\bf Lev Kofman}\\
\vskip .1cm
{Institute for Astronomy, University of Hawaii,
2680 Woodlawn Dr., Honolulu, HI 96822, USA}\footnote{On leave
of absence from  Institute of
Astrophysics and Atmospheric Physics,  Tartu EE-2444, Estonia}
\vskip .2cm
{\bf   Andrei Linde}\\
\vskip .1cm
{Department of Physics, Stanford University, Stanford, CA
94305, USA}\footnote{On leave of absence from Lebedev
Physical Institute, Moscow 117924, Russia}\\
\vskip .3cm
{and} \\
\vskip .3cm
{\bf \ Alexei A. Starobinsky}\\
\vskip .1cm
{Yukawa Institute for Theoretical Physics, Kyoto University,
Uji 611, Japan \\
\vskip .1cm
and Landau Institute for Theoretical Physics,
Kosygina St. 2, Moscow 117334, Russia \\}
\end{center}
\vskip 1.5cm
\begin{quotation}
\centerline{\bf  ABSTRACT}
\vskip .2cm

The  theory of reheating of the Universe after inflation is
developed. We have found that typically at the first stage of reheating the
classical inflaton field $\phi$ rapidly decays into $\phi$-particles or into
other bosons due to a broad parametric resonance. Then these bosons decay into
other particles, which eventually become thermalized. Complete reheating is
possible only in those theories where a single particle $\phi$ can decay into
other particles. This imposes strong constraints on  the structure of
inflationary models, and implies that the inflaton field can be a dark matter
candidate.

\vskip 1.3cm

\centerline{PACS  numbers:\ \ 98.80.Cq, \ 04.62.+v}
\end{quotation}
\end{titlepage}

\vfill
\eject

1.  The theory of reheating of the Universe after inflation is the
most important application of the quantum theory of
particle creation, since almost all matter constituting the
Universe at the subsequent radiation-dominated stage was
created during this process \cite{MyBook}. At the stage of
inflation all
energy was
concentrated in a classical slowly moving inflaton field $\phi$. Soon
after the
end of inflation this field began to oscillate near the minimum of
its
effective potential. Gradually it produced many elementary particles,
they
interacted with each other and came to a state of thermal equilibrium
with some
temperature $T_r$, which was called the reheating temperature.

An elementary theory  of reheating was first
developed in  \cite{1} for the new inflationary scenario.
Independently a theory of reheating in the $R^2$ inflation was constructed in
\cite{st81}.  Various
aspects of this
theory   were further elaborated by many authors, see e.g.
\cite{Dolg}.
 Still, a general scenario of reheating was
absent. In particular, reheating in the chaotic inflation theory
remained
almost unexplored.
The present paper is a short account of our   investigation of
this question  \cite{REH}.  We have found that the process of
reheating typically
consists of  three different stages.  At the first stage, which cannot be
described by the elementary theory of reheating,  the classical coherently
oscillating
inflaton field $\phi$ decays into massive bosons (in particular, into
$\phi$-particles) due to parametric resonance.  In many models the resonance is
very broad, and the process occurs extremely
rapidly (explosively).  Because of the Pauli exclusion principle, there is no
explosive  creation of fermions.
To distinguish this stage from the stage of  particle decay and thermalization,
we will call it {\it pre-heating}. Bosons produced at that stage are far away
from thermal equilibrium and typically have enormously large occupation
numbers. The second stage  is the decay of previously produced  particles. This
stage typically can be described by methods developed in   \cite{1}. However,
these methods should be applied not  to the decay of the original homogeneous
inflaton field, but to the decay of particles and fields produced at the  stage
of explosive reheating. This considerably changes many features of  the
process, including the final value of the reheating temperature.  The
third stage is the stage of thermalization, which can be described by
standard
methods, see e.g. \cite{MyBook,1}; we will not consider it here.
Sometimes this stage
may occur simultaneously with the second one. In our
investigation we
have used the formalism of the
 time-dependent Bogoliubov transformations   to find the density of
created particles, $n_{\vec k}(t)$.
A detailed
description of this theory will be given in \cite{REH}; here we will
outline
our main conclusions using a simple semiclassical
approach.

2.  We will consider a simple chaotic inflation scenario describing the
classical
inflaton scalar field
$\phi$ with the effective potential   $V(\phi) =  \pm {1\over2}
m_\phi^2 \phi^2+{\lambda\over 4}\phi^4$. Minus sign corresponds to
spontaneous symmetry breaking $\phi \to \phi +\sigma$ with generation of a
classical scalar field $\sigma = {m_\phi \over\sqrt\lambda}$.  The field $\phi$
after inflation may decay
into bosons $\chi$ and fermions $\psi$ due to the interaction  terms $- {
1\over2} g^2 \phi^2 \chi^2$ and
 $- h \bar \psi \psi \phi$. Here $\lambda$, $ g$ and  $h$ are
small coupling constants.  In case of  spontaneous symmetry breaking, the term
$- {
1\over2} g^2 \phi^2 \chi^2$ gives rise to the   term  $- g^2 \sigma\phi
\chi^2$. We will assume for simplicity that the bare masses
of the fields $\chi$ and $\psi$ are very small, so that one can write  $ m_\chi
(\phi) =
  g \phi$,  $m_{\psi}(\phi) =  |h\phi|$.

Let us briefly recall the elementary theory of reheating
\cite{MyBook}.  At
$\phi > M_p$, we have a stage of inflation.  This stage is supported
by the
friction-like term $3H\dot\phi$ in the equation of motion for the scalar
field. Here $H\equiv \dot a/a$ is the Hubble parameter, $a(t)$ is the
scale factor of the Universe.
However, with a decrease of the field $\phi$ this term becomes less
and less important, and inflation ends at $\phi {\
\lower-1.2pt\vbox{\hbox{\rlap{$<$}\lower5pt\vbox{\hbox{$\sim$}}}}\ }M_p/2$.
After that the
field $\phi$  begins  oscillating near the minimum of
$V(\phi)$ \cite{f2}. The amplitude of the oscillations  gradually
decreases because of expansion of the
universe, and also because of the energy transfer to particles
created by the
oscillating field.  Elementary
theory of reheating is based on the
 assumption that  the
classical oscillating scalar field $\phi (t)$ can be represented as a
collection of scalar particles at rest. Then the rate of decrease of
the energy of oscillations
coincides with the decay rate  of  $\phi$-particles. The
rates of
the processes $\phi \to \chi\chi$  and $\phi \to  \psi\psi$ (for  $m_\phi \gg
2m_\chi, 2m_\psi$) are given
by
 \begin{equation}\label{7}
  \Gamma ( \phi \to \chi \chi) =  { g^4 \sigma^2\over 8
\pi m_{\phi}}\  , \ \ \ \ \
\Gamma( \phi \to \psi \psi )  =  { h^2 m_{\phi}\over 8 \pi}\ .
 \end{equation}
Reheating
completes when the rate of expansion of the universe given   by the Hubble
constant $H=\sqrt{8\pi \rho\over 3 M^2_p} \sim  t^{-1}$ becomes smaller than
the total decay rate $\Gamma =  \Gamma (\phi \to \chi \chi) + \Gamma
(\phi \to
\psi \psi )$. The reheating temperature can be estimated by
$T_r \simeq 0.1\, \sqrt{\Gamma M_p}$\,.

As we already mentioned, this theory can provide a qualitatively correct
description of particle decay at the last stages of reheating.
Moreover, this theory  is always applicable  if the inflaton field
can decay  into fermions only, with a small coupling constant $h^2 \ll
m_{\phi}/M_p$.
 However,
typically this theory is inapplicable to the description of the first stages of
reheating, which makes the whole process quite different. In what follows we
will develop the theory of the first stages of reheating. We will begin with
the theory of a massive scalar field $\phi$ decaying into particles $\chi$,
then we consider the theory  ${\lambda\over 4} \phi^4$,
and finally we will discuss reheating in the theories with spontaneous symmetry
breaking.

3. We begin with the investigation of the simplest  inflationary
model with the effective potential
${m^2_\phi\over 2}\phi^2$.
Suppose that this field  only interacts
with a light  scalar field $\chi$
 ($m_{\chi} \ll m_{\phi}$) due to the
 term $-{ 1\over2} g^2 \phi^2  \chi^2$.
The equation for quantum fluctuations of the field $\chi$
with the physical momentum $\vec k/a(t)$ has the following form:
\begin{equation}\label{M}
\ddot \chi_k   + 3H \dot \chi_k +  \left({k^2\over a^2(t)}
+ g^2 \Phi^2\, \sin^2(m_{\phi}t) \right) \chi_k = 0 \ ,
\end{equation}
where $k = \sqrt {\vec k^2}$, and $\Phi$ stands for the amplitude of
oscillations of the field $\phi$. As we shall see, the
main contribution to $\chi$-particle
production is given by excitations of the field $\chi$ with
 $k/a \gg m_\phi$, which is much
greater than $H$ at the stage
of oscillations. Therefore, in the first approximation we may neglect
the expansion of the Universe,  taking $a(t)$ as a constant and omitting
the term $3H \dot \chi_k$ in (\ref{M}). Then the equation (\ref{M})
describes an oscillator with a variable
frequency $\Omega_k^2(t)=
 k^2a^{-2} + g^2\Phi^2\, \sin^2(m_{\phi}t) $.
Particle production occurs due to a
nonadiabatic change of this frequency. Equation (\ref{M}) can be
reduced to the well-known  Mathieu equation:
\begin{equation}\label{M1}
\chi_k''   +   \left(A(k) - 2q \cos 2z \right) \chi_k = 0 \ ,
\end{equation}
where   $A(k)
= {k^2 \over m_\phi^2 a^2}+2q$, $q = {g^2\Phi^2\over
4m_\phi^2} $, $z
= m_{\phi}t$, prime denotes differentiation with respect to $z$.
An important property of solutions of the equation (\ref{M1}) is the
existence of an exponential instability $\chi_k \propto \exp
(\mu_k^{(n)}z)$ within the set of resonance bands  of frequencies
$\Delta k^{(n)}$ labeled by an integer index $n$.
This instability corresponds to exponential growth of occupation
numbers of quantum fluctuations
$n_{\vec k}(t) \propto \exp (2\mu_k^{(n)} m_{\phi} t)$
  that may be interpreted as particle
production. The simplest way to analyse this effect is to study the
stability/instability chart of the Mathieu equation,
which is sketched in  Fig. 1. White bands on
this chart correspond to the regions of instability,
the grey bands correspond to regions of stability.
 The  curved lines inside white
bands show the values of the instability parameter $\mu_k$.  The line
$A = 2q$
shows the values of $A$ and $q$ for $k = 0$. All points in the white
regions
above this line correspond to instability for any given $q$. As one
can
see,   near the line $A = 2q$ there are regions in the first,
the second and the  higher instability bands
where the unstable modes grow extremely
rapidly, with $\mu_k \sim 0.2$. We will show analytically in
\cite{REH} that  for $q \gg 1$
 typically
 $\mu_k \sim {\ln 3\over 2\pi}
\approx 0.175$ in the instability bands along the line $A = 2q$,
but its maximal value is ${\ln(1+\sqrt{2}) \over \pi} \approx 0.28$.
Creation of
particles in the regime of a broad  resonance ($q > 1$) with $2\pi \mu_k =
O(1)$ is very different from that in  the usually
considered case of a narrow resonance ($ q \ll 1$),
 where $2\pi \mu_k \ll 1$.
 In particular, it
proceeds during a tiny part of each oscillation of the field $\phi$
when $1-\cos z \sim q^{-1}$ and the induced effective mass of the
field $\chi$ (which is
determined by the condition $m^2_{\chi}= g^2\Phi^2/2$) is less than
$m_{\phi}$.
 As a result, the number of
particles grows exponentially within just a few oscillations of the
field
$\phi$. This leads to an extremely rapid  (explosive)  decay of the
classical
scalar field $\phi$.
This regime occurs only
if  $q {\ \lower-1.2pt\vbox{\hbox{\rlap{$>$}\lower5pt\vbox{\hbox{$\sim$}}}}\ }
\pi^{-1}$, i.e. for $g\Phi {\
\lower-1.2pt\vbox{\hbox{\rlap{$>$}\lower5pt\vbox{\hbox{$\sim$}}}}\ }
m_\phi$, so that $m_\phi \ll gM_p$ is the necessary condition for it.
One can show that a typical energy $E$ of a particle produced at this stage is
determined by
 equation $A-2q \sim \sqrt{q}$, and is given by
 $E  \sim  \sqrt{g m_\phi M_p}$ \cite{REH}.

Creation of $\chi$-particles leads to the two main effects:
transfer of the energy from the homogeneous field $\phi (t)$ to these
particles and generation of the contribution to the effective mass of
the $\phi$ field:  $m^2_{\phi ,eff}=m^2_{\phi}+g^2\langle\chi^2
\rangle_{ren}$.
The last term in the latter expression
quickly becomes larger than
$m^2_{\phi}$. One should take
both these effects into account when calculating backreaction of
created particles on the process.
As a result, the stage of the broad resonance creation ends up within
the short time
$t\sim m_{\phi}^{-1} \ln (m_{\phi}/g^5M_p)$,
when $\Phi^2 \sim
\langle\chi^2\rangle$ and  $q = {g^2\Phi^2\over
4m_{\phi ,eff}^2}$
becomes smaller than $1$.
At this time the energy density of produced particles
$\sim E^2 \langle\chi^2\rangle \sim g m_\phi M_p \Phi^2$ is of the same
order as the original energy density
$\sim {m_\phi^2} M_p^2$ of the scalar field
$\phi$ at the end of inflation. This gives the amplitude of
oscillations at the end of the stage of the broad resonance particle creation:
$\Phi^2 \sim \langle\chi^2\rangle \sim
g^{-1} m_\phi M_p \ll M_p^2$.
Since $E\gg m_{\phi}$, the effective equation of state of the whole
system becomes $p\approx \varepsilon /3$. Thus, explosive creation
practically eliminates a prolonged intermediate matter-dominated stage
after the end of inflation which was thought to be characteristic
to many inflationary models.
However, this does not mean that the process of reheating has been completed.
Instead of $\chi$-particles in the thermal equilibrium with
 a typical energy
 $E \sim T \sim (mM_p)^{1/2}$, one has particles with a much
smaller energy $\sim  (g m_\phi M_p)^{1/2}$,
but with extremely large
mean occupation numbers  $n_k \sim g^{-2} \gg 1$.

After that the Universe expands as $a(t)\propto \sqrt t$, and
the scalar field $\phi$ continues its decay in the regime of the narrow
resonance creation $q\approx {\Phi^2\over 4 \langle\chi^2\rangle}
\ll 1$. As a result,
$\phi$ decreases rather slowly, $\phi \propto t^{-3/4}$.
This regime is very important  because  it makes the energy of the
$\phi$ field much smaller than that of the $\chi$-particles.
One can show that the decay finally stops when the amplitude of
oscillations $\Phi$ becomes smaller than $g^{-1} m_\phi$ \cite{REH}.
This happens at the moment $t\sim  m_{\phi}^{-1}  (gM_p/m_{\phi})^{1/3}$
(in the case   $m < g^7 M_p$ decay ends somewhat later,
in the perturbative regime).
The physical reason why the decay  stops is rather general: decay of
the particles $\phi$ in our model occurs due to its interaction with another
$\phi$-particle (interaction term is quadratic in $\phi$ and in $\chi$). When
the  field $\phi$ (or the number of $\phi$-particles) becomes
small, this process
is inefficient.  The scalar field can decay completely only if a
single  scalar $\phi$-particle can decay into other  particles, due
to the processes
$\phi \to \chi \chi$ or $\phi \to \psi \psi$, see eq. (\ref{7}). If
 there is
 no spontaneous symmetry breaking and no interactions with fermions
in our model, such  processes are impossible.

At later stages the energy of oscillations of the inflaton field
decreases as $a^{-3}(t)$, i.e. more slowly than the decrease of
energy of hot ultrarelativistic matter $\propto a^{-4}(t)$. Therefore, the
relative contribution of the field $\phi(t)$ to the total energy density
of the Universe
rapidly grows. This   gives rise to an unexpected possibility that
the inflaton field  by
itself, or other scalar fields can be
cold dark matter candidates, {\it even if they strongly interact with each
other}. However, this possibility requires
a certain degree of fine tuning; a more immediate application of our result is
that it allows one to rule out a wide class of inflationary models which do not
contain interaction terms of the type of $g^2\sigma\phi\chi^2$ or
$h\phi\bar\psi\psi$.

4. So far we have not considered  the term ${\lambda \over 4} \phi^4$
in the effective potential. Meanwhile this term leads to production
of $\phi$-particles, which in some cases appears to be the leading
effect.
 Let us  study the $\phi$-particle production in the theory
with $V(\phi)  =
{m^2_{\phi}\over 2} \phi^2 + {\lambda\over 4}\phi^4$ with $m^2_{\phi}
\ll \lambda  M_p^2$. In this case the effective potential
of the field $\phi$ soon after the end of inflation at
$\phi \sim M_p$ is dominated by the term
${\lambda\over 4} \phi^4$.  Oscillations  of the field $\phi$ in this
theory
are not sinusoidal, they are
given by elliptic functions, but with a good accuracy one can write
$\phi(t)
\sim \Phi \sin (c\sqrt \lambda \int \Phi dt)$, where
$c={\Gamma^2(3/4)\over \sqrt \pi} \approx 0.85$.   The Universe at
that time expands as at the
radiation-dominated stage: $a(t)\propto \sqrt t$. If one neglects
the feedback of created $\phi$-particles on the homogeneous field
$\phi (t)$, then its amplitude $\Phi (t) \propto a^{-1}(t)$, so that $a\Phi
=const$.
Using a conformal time $\eta$, exact equation for quantum fluctuations
 $\delta \phi$
 of the field $\phi$ can be reduced to the Lame equation. The results remain
essentially the same if we use an  approximate equation
\begin{equation}\label{lam1}
{d^2(\delta\phi_k)\over d\eta^2}   +   {\Bigl[{k^2} +
3\lambda a^2\Phi^2\, \sin^2 (c\sqrt\lambda a\Phi \eta)\Bigr]}
\delta\phi_k = 0 \ ,~~~\eta =\int {dt\over a(t)}={2t\over a(t)}\, ,
\end{equation}
which leads to the Mathieu equation with $A =
{k^2\over c^2\lambda a^2\Phi^2} +
{3\over 2c^2} \approx  {k^2\over c^2\lambda a^2\Phi^2} + 2.08$, and
$q = {3\over 4c^2} \approx 1.04$.   Looking at the instability chart, we see
that the
resonance occurs in the second band, for $k^2 \sim 3\lambda a^2\Phi^2$. The
maximal value of the coefficient $\mu_k$ in   this band for $q \sim 1$
approximately equals to $0.07$. As long as the backreaction of created
particles is small, expansion of the Universe does not shift fluctuations away
from the resonance band, and the
number of produced particles grows  as  $\exp (2c\mu_k\sqrt\lambda a\Phi \eta)
\sim
\exp ({\sqrt\lambda\Phi t\over 5})$.

After the time interval $\sim M_p^{-1}\lambda^{-1/2}|\ln \lambda|$,
  backreaction of created particles  becomes significant. The growth of the
fluctuations
$\langle\phi^2\rangle$ gives rise to a   contribution
$3\lambda \langle\phi^2\rangle$ to the effective mass squared of the field
$\phi$, both in the equation for $\phi (t)$ and in Eq. (\ref{lam1}) for
inhomogeneous modes.
The stage of explosive reheating ends when $\langle\phi^2\rangle$ becomes
greater than $\Phi^2$. After that, $\Phi^2 \ll
\langle\phi^2\rangle$ and
the effective frequency of oscillations is determined by the
term $\sqrt{3\lambda \langle\phi^2\rangle}$.
The corresponding process is
 described by Eq. (4) with  $A(k) = 1 + 2q + {k^2
\over
3\lambda a^2\langle\phi^2\rangle}$, $q = {\Phi^2\over 4
\langle\phi^2\rangle}$. In this regime $q \ll 1$, and particle creation occurs
in the  narrow resonance regime in the second band with $A \approx 4$.  Decay
of the field in this regime is extremely slow: the amplitude $\Phi$ decreases
only by a factor
 $t^{1/12}$ faster that  it would decrease without any decay, due to the
expansion of the Universe only, i.e., $\Phi \propto t^{-7/12}$ \cite{REH}.
Reheating stops altogether when the presence of non-zero mass
$m_{\phi}$ though still small as compared to $\sqrt{3\lambda
\langle\phi^2\rangle}$
appears enough for the expansion of the Universe to drive
a mode away
from the narrow resonance. It happens when the amplitude $\Phi$ drops
up to a value $\sim m_{\phi}/\sqrt \lambda$.

In addition to this process, the field $\phi$ may decay  to
$\chi$-particles.
This is the leading process for    $g^2\gg \lambda$.
The equation for $\chi_k$ quanta has  the same form as eq.
(\ref{lam1})
with the obvious change $\lambda \to g^2/3$.
Initially  parametric resonance is broad. The values of the parameter
$\mu_k$
along the line $A = 2q$ do not change monotonically, but typically
for $q \gg
1$ they are 3 to 4 times greater than the parameter $\mu_k$ for the
decay of
the field $\phi$ into its own quanta. Therefore, this pre-heating
process is very
efficient. It ends at the moment $t\sim M_p^{-1}\lambda^{-1/2}
\ln (\lambda /g^{10})$ when $\Phi^2 \sim \langle \chi^2
\rangle \sim g^{-1}\sqrt \lambda M_p^2$. The typical energy of created
$\chi$-particles is $E \sim (g^2\lambda)^{1/4}M_p$. The following
evolution is essentially the same as that described in Sec. 3.

5. Finally, let us consider the case with symmetry breaking.  In the
beginning, when the amplitude of oscillations is much greater than
$\sigma$, the theory of  decay of the inflaton field is the same as in the case
considered above. The most important part of  pre-heating occurs at this stage.
When the amplitude of the oscillations becomes smaller than
$m_\phi/\sqrt\lambda$ and the field begins oscillating near the minimum of the
effective potential at $\phi = \sigma$, particle production due to
the narrow parametric
resonance typically becomes  very weak.
The main reason for this is related to the backreaction of
particles created at the
preceding stage of pre-heating on the rate of expansion of the universe and on
 the shape of the effective potential \cite{REH}. However, importance of
spontaneous symmetry
breaking for the theory of reheating should not be underestimated, since it
gives rise to the interaction term   $g^2\sigma\phi\chi^2$ which is linear in
$\phi$. Such terms are necessary for a complete decay of the inflaton field in
accordance with the perturbation theory (\ref{7}).

6. In this paper we discussed the process of reheating of the universe   in
various inflationary models. We have found that  decay of the inflaton field
typically begins with a stage of explosive production of particles at a stage
of a broad parametric resonance.
Later the resonance becomes narrow, and
finally this stage of
decay finishes altogether. Interactions of  particles produced at this
stage, their decay into other particles and subsequent thermalization typically
require  much more time that the stage of pre-heating, since these processes
are suppressed by the small
values of coupling constants. The corresponding processes in
many cases can be described by the elementary theory of reheating. However,
this  theory  should be applied not to the decay of the
original large and homogeneous oscillating inflaton field, but to the decay  of
particles produced at the stage of pre-heating, as well as to the decay of
small remnants of the  classical   inflaton field. This makes a lot
of difference, since typically coupling constants of interaction of the
inflaton field with matter are extremely small, whereas coupling constants
involved in the decay of  other bosons  can be much greater. As a result,
the reheating temperature can be much higher than the typical temperature $T_r
{\ \lower-1.2pt\vbox{\hbox{\rlap{$<$}\lower5pt\vbox{\hbox{$\sim$}}}}\ } 10^9$
GeV
which could be obtained
neglecting the stage of parametric resonance \cite{REH}.
On the other hand, such processes as baryon creation after inflation occur best
of all outside  the state of thermal equilibrium. Therefore, the stage
of   pre-heating  may play an extremely important role in our cosmological
scenario.  Another  consequence of the resonance effects is an almost
instantaneous change of equation of state from the vacuum-like one to the
equation of state of relativistic matter. This leads to
suppression of  the number of  primordial black holes  which could be produced
after inflation.

L.K. was supported in part by the Canadian Institute for Advance
Research
cosmology program, and CITA.   A.L.  was supported in part  by NSF
grant
PHY-8612280.
A.S. is grateful to Profs. Y. Nagaoka and J. Yokoyama for their
hospitality at the Yukawa Institute for Theoretical Physics, Kyoto
University.   A.S.  was supported in part  by the Russian Foundation
for
Basic Research, Project Code 93-02-3631, and by Russian Research
Project ``Cosmomicrophysics''.

\

{\large \bf Figure Caption:}

{ \bf Fig. 1.} \ The sketch of the stability/instability
chart of the canonical Mathieu equation (3).


\begin{thebibliography}{999}
\bibitem{f2} {Some authors discussed a possibility   to add
 additional terms of the type of $\Gamma \dot\phi$ to
the equation of motion in order
to describe energy transfer due to particle creation. There was a
hope that such terms
may lead to a longer stage of inflation. We have found that this
prescription
does not correctly describe particle creation at the stage preceding
a stage of
oscillations of the scalar field, and that it is not possible to
support inflation by the term $\Gamma \dot\phi$ \cite{REH}. }
\bibitem{f3}{It
is interesting that the rapid decay of the field $\phi$ can occur
even if the
field $\chi$ has a bare mass $m_\chi(0)$ much greater than the
frequency of
oscillations $m_\phi$. This rather unexpected regime is possible if
the
amplitude of oscillations is sufficiently large, $g\Phi \gg
m_{\chi}^2/ m_\phi$.}
\end{thebibliography}
\end{document}